
%
%
%
%
\ifx\epsffile\undefined\message{(FIGURES WILL BE IGNORED)}
\def\insertfig#1{}
\else\message{(FIGURES WILL BE INCLUDED)}
\def\insertfig#1{{{
\midinsert\centerline{\epsfxsize=\hsize
\epsffile{#1}}\bigskip\bigskip\bigskip\bigskip\endinsert}}}
\fi

\input harvmac
%
%
%
%
\ifx\answ\bigans
\else
\output={
  \almostshipout{\leftline{\vbox{\pagebody\makefootline}}}\advancepageno
}
\fi
%
%
%

%
%

%
%
\def\UCSD#1#2{\noindent#1\hfill #2%
\bigskip\supereject\global\hsize=\hsbody%
\footline={\hss\tenrm\folio\hss}}
%
%
\def\abstract#1{\centerline{\bf Abstract}\nobreak\medskip\nobreak\par #1}
%
%
%
%
\edef\tfontsize{ scaled\magstep3}
 \tfontsize  \tfontsize
 \tfontsize \font\titlei=cmmi10 \tfontsize
\font\titleis=cmmi7 \tfontsize \font\titleiss=cmmi5 \tfontsize
\font\titlesy=cmsy10 \tfontsize \font\titlesys=cmsy7 \tfontsize
\font\titlesyss=cmsy5 \tfontsize  \tfontsize
\skewchar\titlei='177 \skewchar\titleis='177 \skewchar\titleiss='177
\skewchar\titlesy='60 \skewchar\titlesys='60 \skewchar\titlesyss='60
%
%
%
%
%
\def\inv{^{\raise.15ex\hbox{${\scriptscriptstyle -}$}\kern-.05em 1}}
\def\lbar{{\lower.35ex\hbox{$\mathchar'26$}\mkern-10mu\lambda}} 

%
%
%
%
\def\dsl{\,\raise.15ex\hbox{/}\mkern-13.5mu D} 
\def\delsl{\raise.15ex\hbox{/}\kern-.57em\partial}
\def\Ksl{\hbox{/\kern-.6000em\rm K}}
\def\Asl{\hbox{/\kern-.6500em \rm A}}
\def\Dsl{\hbox{/\kern-.6000em\rm D}} 
\def\Qsl{\hbox{/\kern-.6000em\rm Q}}
\def\gradsl{\hbox{/\kern-.6500em$\nabla$}}
%
%
\def\lspace{\ifx\answ\bigans{}\else\qquad\fi}
\def\lbspace{\ifx\answ\bigans{}\else\hskip-.2in\fi} 
%
%
\def\boxeqn#1{\vcenter{\vbox{\hrule\hbox{\vrule\kern3pt\vbox{\kern3pt
        \hbox{${\displaystyle #1}$}\kern3pt}\kern3pt\vrule}\hrule}}}
%
%
\def\mbox#1#2{\vcenter{\hrule \hbox{\vrule height#2in
\kern#1in \vrule} \hrule}}
%
%
%
%
   \def\CD{{\cal D}}
   
   \def\CL{{\cal L}}
  \def\CO{{\cal O}}

%
%
%
%
%

%

\def\abs#1{\left| #1\right|}

\def\darr#1{\raise1.5ex\hbox{$\leftrightarrow$}\mkern-16.5mu #1}

%
%
\def\frac#1#2{{\textstyle{#1\over #2}}} 
%
%
%
%

\def\Tr{\mathop{\rm Tr}}

%
%
%
%

%
%
\def\ltap{\ \raise.3ex\hbox{$<$\kern-.75em\lower1ex\hbox{$\sim$}}\ }
\def\gtap{\ \raise.3ex\hbox{$>$\kern-.75em\lower1ex\hbox{$\sim$}}\ }
\def\gl{\ \raise.5ex\hbox{$>$}\kern-.8em\lower.5ex\hbox{$<$}\ }
\def\roughly#1{\raise.3ex\hbox{$#1$\kern-.75em\lower1ex\hbox{$\sim$}}}
%
%
        \def\etc{\hbox{\it etc.}}

\def\np#1#2#3{Nucl. Phys. B{#1} (#2) #3}
\def\pl#1#2#3{Phys. Lett. {#1}B (#2) #3}
\def\prl#1#2#3{Phys. Rev. Lett. {#1} (#2) #3}
\def\physrev#1#2#3{Phys. Rev. {#1} (#2) #3}

\relax

\noblackbox

\centerline{{\titlefont{Semiclassical Meson-Baryon Dynamics}}}
\medskip
\centerline{{\titlefont{from Large-$N_c$ QCD}}}
\bigskip
\centerline{Aneesh V.~Manohar}
\smallskip
\centerline{{\it Department of Physics, University of California at
San Diego,}}
\centerline{{\it 9500 Gilman Drive, La Jolla, CA 92093}}
\vfill
\abstract{The large-$N_c$ limit of the meson-baryon effective Lagrangian
is shown to reduce to a semiclassical field theory. A chiral bag
structure emerges naturally in the $N_c\rightarrow \infty$ limit. A
possible connection between the chiral bag picture and the Skyrme model
is discussed.  The classical meson-baryon theory is used to reproduce
the $M_\pi^3$ non-analytic correction to the baryon mass obtained
previously as a loop correction in chiral perturbation theory. }
\vfill
\UCSD{\vbox{\hbox{UCSD/PTH 94--14}\hbox{hep-ph/9407211}}}{July 1994}

Quantum chromodynamics has an expansion in $1/N_c$, where $N_c$ is the
number of colors. This expansion was originally proposed by 't~Hooft
\ref\thooft{G.~'t~Hooft, \np {72} {1974} {461}, \np {75} {1974} {461}},
who used it to solve for the exact meson spectrum in 1+1 dimensions. The
baryon sector of QCD in the large-$N_c$ limit was investigated by Witten
\ref\witten{E.~Witten, \np {160} {1979} {57}}. Recently, progress has
been made in calculating baryon properties in a systematic expansion in
$1/N_c$ \ref\dm{R.~Dashen and A.V.~Manohar, \pl {315} {1993} {425,
438}}\ref\ej{E.~Jenkins, \pl {315} {1993} {431, 441, 447}}. The baryon
sector of QCD has a contracted $SU(6)$ spin-flavor symmetry in the
$N_c\rightarrow \infty$ limit \ref\gs{J.-L.~Gervais and B.~Sakita,
\prl{52} {1984} {527}, \physrev {D30} {1984} {1795}}\dm. The form of the
spin-flavor symmetry violation away from $N_c=\infty$ can be classified
in terms of $1/N_c$ suppressed operators. Results can be obtained for
baryons to all orders in $SU(3)$ breaking \ref\djm{R.~Dashen, E.~Jenkins
and A.V.~Manohar, \physrev {D49}{1994}{4713}\semi E.~Jenkins and
A.V.~Manohar, {\it Baryon Magnetic Moments in the $1/N_c$ Expansion},
UCSD/PTH 94--10, {\tt [hep-ph/9405431]}}. A complementary approach to
the one developed in refs.~\dm\ej\djm\ has been used in
refs.~\ref\cgo{C.~Carone, H.~Georgi, and  S.~Osofsky, \pl {322} {1994}
{227}}\ref\lmr{M.~Luty and J.~March-Russell, {\it Baryons from Quarks in
the $1/N_c$ Expansion}, LBL preprint LBL--34778 {\tt
[hep-ph/9310369]}}\ref\cgkm{C.D.~Carone, H.~Georgi, L.~Kaplan, and
D.~Morin, {\it Decay of $\ell=1$ Baryons --- Quark Model versus
Large-$N_c$}, HUTP--94/A008, {\tt [hep-ph/9406227]}}. The results
obtained using the $1/N_c$ expansion in QCD are closely related to
earlier results obtained using the Skyrme model \ref\anw{G.S. Adkins,
C.R. Nappi, and E. Witten, \np {228} {1983} {552}} \ref\mattis{M.P.
Mattis and M. Mukerjee, \prl {61} {1988} 1344\semi M.P. Mattis and E.
Braaten, \physrev{D39} {1989} 2737}\ref\avm{A.V. Manohar, \np {248}
{1984} {19}}.

Meson-baryon interactions at low momentum transfer can be described by
an effective Lagrangian. In this paper, we will show how the effective
Lagrangian in the $N_c\rightarrow \infty$ limit can be treated as a
classical field theory which is similar to the chiral bag picture of
hadrons \ref\chiralbag{A.~Chodos and C.~Thorn, \physrev{D12} {1975}
{2733}\semi M.~Rho, A.S.~Goldhaber, and G.E.~Brown, \prl{51} {1983}
{74}}. This connection was also discussed previously by Gervais and
Sakita \gs. The classical theory will be used to reproduce some known
results obtained previously using chiral perturbation theory. In
particular, we will show how the classical field theory sums all the
non-analytic chiral corrections which are of leading order in $1/N_c$.

The $N_c$ counting rules imply that in general, three-meson couplings
are of order $1/\sqrt{N_c}$, four meson couplings are of order $1/N_c$,
\etc\ Thus the meson self-couplings are described by an effective
Lagrangian of the form
\eqn\genmeson{
\CL_M = N_c\ \CL^{(M)}\left(\left\{\frac{M_i}{\sqrt N_c}\right\}\right),
}
where $\{M_i\}$ is the set of meson fields. Meson-baryon couplings are
of order $\sqrt N_c$, two-meson--baryon couplings are of order one,
\etc, so the meson-baryon interaction Lagrangian can be written as
\eqn\genint{
\CL_B = N_c\ \CL^{(B)}\left(\left\{\frac{M_i}{\sqrt N_c}\right\}, B \right),
}
where $B$ is the baryon field. In the large $N_c$ limit, the baryon is
infinitely heavy and can be treated as a static source localized at the
origin. The contracted $SU(4)$ symmetry of large-$N_c$ QCD implies that
there is an infinite degenerate baryon tower of states with $I=J=1/2,\
3/2,\ \ldots$ \gs\dm. It is more convenient to represent this infinite
tower of baryon states by a collective coordinate $A$ from which the
$I=J$ baryons can be obtained by projection \anw\avm. Since the baryon
mass splittings are order $1/N_c$ \ej, the collective coordinate $A$
does not evolve with time at leading order in $1/N_c$. Thus the
large-$N_c$ baryons can be replaced by a classical source with
collective coordinate $A$ located at $\vec x = \vec x_0$. The
interaction Lagrangian eq.~\genint\ can be written as
\eqn\geninti{
\CL_A = N_c\ \CL^{(A)}\left(\left\{\frac{M_i}{\sqrt N_c}\right\}, A, \vec x_0
\right), }
where $A$ is a time independent classical field. The total Lagrangian for
meson-baryon interactions is
\eqn\ltotal{
\CL = N_c\left[ \CL^{(M)}\left(\left\{M_i\right\}\right)
+ \CL^{(A)}\left(\left\{M_i\right\}, A, \vec x_0 \right) \right]=N_c\ \overline
\CL\left(\left\{M_i\right\}, A, \vec x_0 \right),
}
on rescaling the meson fields by $\sqrt N_c$. The functional integral is
performed over the meson fields $M_i$ (but not over the time-independent
collective coordinate $A$),
\eqn\funcint{
Z = \int \left\{ \CD M_i\right\}\ e^{i {N_c} \overline S/\hbar},
}
where
\eqn\sdef{
\overline S = \int d^4 x\ \overline \CL\left(\left\{M_i\right\}, A, \vec
x_0 \right).
}
The functional integral in eq.~\funcint\ needs to be regulated. The effective
meson-baryon Lagrangian is valid for momentum scales small compared to
$\Lambda_\chi\sim1$~GeV. The functional integral eq.~\funcint\ can be
regulated by using $\Lambda_\chi$ as a cutoff. The details of the cutoff
procedure are unimportant; what is important is that the cutoff
$\Lambda_\chi$ is $N_c$-independent so that there is no hidden $N_c$
dependence in the functional measure. There is an overall factor of
$N_c/\hbar$ in front of the entire action, so the $1/N_c$ expansion is
equivalent to the semiclassical expansion in powers of $\hbar$.\foot{The
connection between the sum of large-$N_c$ diagrams and the semiclassical
approximation was proven by Arnold and Mattis for a certain class of
diagrams \ref\arnold{P.~Arnold and M.P.~Mattis, \prl{65} {1990}
{831}}.}\ The leading term is the solution of the classical equations of
motion of eq.~\ltotal. This produces a baryon source $A$ with a
classical meson cloud, and is closely related to the chiral bag model.
The quantum corrections are obtained by performing a semiclassical
expansion about the classical meson background, and including
time-dependence in the baryon collective coordinate $A$. Note that the
infinite tower of baryon states is crucial for a classical representation of
baryons. A single $I=J=1/2$ state cannot be treated as a classical object.

We will now study a specific example of the meson-baryon effective
Lagrangian in more detail. Consider the interaction of pions with the
$I=J=1/2, 3/2, \ldots$ baryon tower to lowest order in the chiral
expansion for two light flavors, and in the isospin limit $m_u=m_d=m_q$.
The pion chiral Lagrangian is
\eqn\pionlag{
\CL^{(M)} = {f_\pi^2\over 4}\ \Tr\, \partial_\mu U \partial^\mu U^\dagger +
{M^2_\pi
\,f^2_\pi \over 4}\ \Tr\, \left[U + U^\dagger\right],
}
where $M_\pi$ is the pion mass, $U$ is the exponential of the pion field
\eqn\defu{
U = e^{2 i \pi/f_\pi},
}
and $f_\pi\approx 93$~MeV is the pion decay constant. The lowest order
pion-baryon coupling is
\eqn\pionb{
\CL^{(A)} =  3 N_c\ g\ \delta\left(\vec x\right)\ X^{ia}\ A^{ia}\left(x\right),
}
where $X^{ia}$ is the baryon axial current in the large-$N_c$ limit,
\eqn\defx{
X^{ia} = \Tr\, A \tau^a A^{-1} \tau^i ,
}
$A^{ia}$ is proportional to the pion axial current,
\eqn\defa{
A^{ia}= {i \over 4}\ \Tr\, T^a\left( U \nabla^i U^\dagger - U^\dagger \nabla^i
U\right),
}
and the static baryon is located at $\vec x=0$.  The baryon axial vector
current $g_A$ is of order $N_c$, so we have written $g_A\equiv N_c g$.
The factor of three in eq.~\pionb\ ensures that the axial coupling of
the $I=J=1/2$ nucleon is $g_A$ when we convert from the collective
coordinate basis to the $I,J$ basis \anw. Expanding the interaction term
eq.~\pionb\ to first order in the pion field gives the pion-baryon vertex
\eqn\pbvertex{
 {3 N_c g\over 2 f_\pi}\ X^{ia}\ \nabla^i \pi^a =  {3g_A\over 2 f_\pi}\ X^{ia}\
\nabla^i \pi^a.
}

The leading non-analytic correction to the baryon mass due to
pion-baryon interactions is from the graph of \fig\oneloop{One loop
correction to the baryon mass.}, and can be computed in chiral
perturbation theory using the interaction vertex eq.~\pbvertex.  The
mass shift for the spin-1/2 nucleon from \oneloop\ is a standard
computation using baryon chiral perturbation theory with a static baryon
\ref\staticchiral{E.~Jenkins and A.V.~Manohar, \pl{255} {1991} {558},
\pl{259} {1991} {353}} including intermediate nucleon and delta states
\ref\ejmasses{E.~Jenkins, \np{368} {1992} {190}}\ej, with $g_{\pi NN}/g_{\pi
N\Delta}$ in the ratio given by $SU(4)$ spin-flavor symmetry,
\eqn\mshift{
\delta m = - \left( {3\over2} + 3\right){g_A^2\, M_\pi^3\over 16 \pi f_\pi^2}
=- {9g^2\, N_c^2\,  M_\pi^3\over 32 \pi f_\pi^2}.
}
The two coefficients in eq.~\mshift\ are the intermediate nucleon and delta
contributions, respectively. Note that $\delta m$ is of order $N_c$,
since $f_\pi\propto \sqrt N_c$, and $g$ and $M_\pi$ are of order one.
The mass-shift in eq.~\mshift\ is non-analytic in the quark mass $m_q$
($\delta m \propto m_q^{3/2}$), since $M_\pi\propto m_q^{1/2}$.

The general form of the baryon masses in the $1/N_c$ expansion is \ej
\eqn\bmass{
m = N_c\ m_0 + m_1 + m_2\ {J^2\over N_c} + \CO\left({1\over N_c^2}\right),
}
where $m_i$ are arbitrary functions of the quark mass $m_q$. Eq.~\mshift\ is an
order $N_c$ loop correction which does not violate the general form for
the baryon masses given in eq.~\bmass. It produces a $m_q^{3/2}$
non-analytic correction that is the same for the entire baryon tower,
and so can be absorbed into $m_0$. The numerical magnitude of the
$m_q^{3/2}$ correction due to pion loops is small. However, the
corresponding $m_s^{3/2}$ correction due to kaon loops is naively of
order 1~GeV, which is comparable to $m_0$ and cannot be treated as a
small perturbation. It is therefore useful to have a calculational
method that directly sums all the order $N_c$ loop corrections, so that
the remaining corrections are suppressed by $1/N_c$, and are small. This
can be done using the semiclassical method outlined above.

We now compute the baryon mass correction due to pion interactions using the
semiclassical approximation, which sums all the order $N_c$ terms. The
classical field equations are solved for a fixed value of the baryon
collective coordinate $A$. It is convenient to choose $A=1$, so that
$X^{ia}$ in eq.~\defx\ is $\delta^{ia}$. The solutions for other values
of $A$ can be obtained trivially by an isospin rotation on the solution
for $A=1$. The Lagrangian for $A=1$ is
\eqn\aonelag{
\CL^{(M)} = {f_\pi^2\over 4}\ \Tr\, \partial_\mu U \partial^\mu U^\dagger +
{M^2_\pi\, f^2_\pi \over 4}\ \Tr\, \left[U + U^\dagger\right]
+ 3 g_A\ \delta\left(\vec x\right)\ \delta^{ia}\ A^{ia},
}
which is of order $N_c$ since $f_\pi\propto \sqrt N_c$ and $g_A\propto
N_c$. The classical equations of motion are those for a pion field with
a $\delta$-function source at the origin. It is clear from the form of
the source term that the pion field has the hedgehog form,
\eqn\hedgehog{
U = e^{i \vec\tau \cdot \hat x F\left(r\right)},
}
for some function $F(r)$. The equations of motion for $F\left(r\right)$ for a
static classical pion field are obtained by minimizing the energy functional
\eqn\efunc{\eqalign{
E &= 2 \pi f_\pi^2 \int_0^\infty dr\ r^2 \left[ \left({\partial F\over\partial
r}\right)^2 + 2\, {\sin^2 F\over r^2} + 2 M_\pi^2\,\left(1-\cos
F\right)\right]\cr
&\qquad\qquad- {3\over 2}g_A  \int d^3 \vec x\ J\left(\abs{\vec x} \right)
\left[{\partial F\over\partial r} + 2\, {\sin F\cos F\over r}\right],
}}
where we have smeared out the baryon source $\delta(x)$ into
$J(\abs{\vec x})$ (for reasons which will become clear). The source is
normalized to unit baryon number,
\eqn\jsum{
\int d^3 \vec x\ J\left(\abs{\vec x }\right) = 1.
}
The equation of motion obtained by varying eq.~\efunc\ is
\eqn\eqmotion{
{\partial\over\partial r}\left(r^2 {\partial F\over \partial r}\right)
- \sin 2F - M_\pi^2\, r^2 \sin F = {3 g_A\over 2 f_\pi^2}
\left[ {\partial\over\partial r}\left(r^2 J(r)\right) - 2 J(r)\, r \cos
2 F\right].
}
This equation was also considered by Gervais and Sakita in their study
of the chiral bag \gs. It is convenient to choose the baryon source
$J(r)$ to be
\eqn\jsource{
J\left(r\right) = \cases{ J_0 &\qquad $r \le R_0$,\cr
0&\qquad $r > R_0$.\cr}
}
with
\eqn\jzero{
{4\over 3} \pi\, R_0^3\, J_0 =1,
}
so that the $\delta$-function is recovered in the limit $R_0\rightarrow
0$. Far away from the baryon source, the pion fields are weak, and
eq.~\eqmotion\ can be approximated by keeping the lowest order terms in $F$,
\eqn\eqmF{
r^2 {d^2 F\over dr^2} + 2 r {dF\over dr} - (2 + M_\pi^2\, r^2) F = {3g_A\over
2f_\pi^2} \ r^2 {d J\over dr},
}
and setting $J=0$. The solution that is regular at infinity is
\eqn\Foutside{
F\left(r\right) = a\ k_1\left(M_\pi r\right),
}
where $a$ is an overall normalization constant, and $k_1$ is a spherical Bessel
function. The asymptotic form of eq.~\Foutside\ is
$$
F(r) \sim {e^{-M_\pi r}\over r},
$$
which produces an exponentially decaying pion tail at infinity.

The chiral Lagrangian is an expansion in derivatives over the chiral symmetry
breaking scale $\Lambda_\chi\sim 1$~GeV, which is held fixed as $N_c\rightarrow
\infty$. The chiral Lagrangian can be used for computing processes in which the
momentum transfer $p\ll \Lambda_\chi$. The $\delta$-function baryon
source is regarded as ``pointlike'' on the scale of the typical momentum
transfer $M_\pi$, but slowly varying on the scale $\Lambda_\chi$. In
other words, one considers the baryon source to be smeared over a region
$R_0$, with $M_\pi \ll R_0^{-1} \ll \Lambda_\chi$. $R_0^{-1}$ is a
cutoff on the pion-baryon interaction vertex. At distances much shorter
than $R_0$, the pion-baryon interaction is no longer pointlike. Chiral
perturbation theory corresponds to computing in the limit $M_\pi \ll
R_0^{-1} \ll \Lambda_\chi$, and treating $R_0^{-1}$ as a cutoff which is
removed at the end of the calculation.

In the limit $M_\pi \ll R_0^{-1} \ll \Lambda_\chi$, the function $F$
is obtained for all values of $r$ using the linear approximation to
eq.~\eqmotion. The solution of eq.~\eqmF\ is easy to obtain since the
inhomogeneous term on the right hand side is non-zero only at $r=R_0$.
The answer is
\eqn\Fsoln{
F\left(r\right) = \cases{ a\ k_1\left(M_\pi r\right)&\qquad $r>R_0$,\cr
b\ i_1\left(M_\pi r\right)&\qquad $r<R_0$,\cr}
}
where $a$ and $b$ are determined from the boundary condition at $r=R_0$ to be
\eqn\absoln{\eqalign{
a &= {3 g_A\over 2 f_\pi^2}\ J_0\, M_\pi\, R_0^2\ i_1(M_\pi R_0),\cr
b &= {3 g_A\over 2 f_\pi^2}\ J_0\, M_\pi\, R_0^2\ k_1(M_\pi R_0).\cr
}}

The mass shift of the baryon due to the pion cloud is obtained by substituting
eqs.~\Fsoln\ and \absoln\ into the energy functional, eq.~\efunc. The resulting
expression includes all corrections of order $N_c$ to the baryon energy,
including any dependence on $M_\pi$ that is of order $N_c$,
\eqn\mfinal{
\delta m = {81\, g_A^2\over 64 \pi f_\pi^2\, M_\pi^3\, R_0^6} \left(1+M
R_0\right)
\left[\left(1-M R_0\right) - \left(1+M R_0\right)e^{-2M R_0}\right].
}
Expanding in powers of the cutoff $R_0^{-1}$, one obtains
\eqn\massshift{
\delta m = -{27\, g_A^2 \over 32 \pi f_\pi^2\, R_0^3} + {27\, g_A^2\,
M_\pi^2\over 80 \pi f_\pi^2\, R_0} - {9\, g_A^2\, M_\pi^3\over 32 \pi
f_\pi^2}+\ldots\ .
}
The first term in eq.~\massshift\ is a cutoff dependent shift in the baryon
mass that is independent of $M_\pi$, and can be asorbed into the chiral
invariant baryon mass term in the chiral Lagrangian. The second term in
eq.~\massshift\ is a cutoff dependent term that is of order $M_\pi^2$,
and hence is analytic in the quark masses. It can be reabsorbed into the
baryon mass term in the chiral Lagrangian that is linear in the quark
mass matrix. There is no term of order $M_\pi/R_0^2$ in eq.~\massshift.
Such a term would require a non-analytic counterterm in the chiral
Lagrangian, which does not exist. Note that there are also no terms in
eq.~\massshift\ which are proportional to inverse powers of $M_\pi$. The
third term in eq.~\massshift\ is of order $M_\pi^3 \propto m_q^{3/2}$
and is non-analytic in the quark masses, so it cannot be absorbed into a
local Lagrangian counterterm. The third term is finite, independent of
the cutoff $R_0$, and reproduces correctly the known result eq.~\mshift\
obtained using chiral perturbation theory.\foot{The first two terms are
not present in the chiral perturbation theory calculation \ej\ejmasses\
because they vanish in dimensional regularization.}\ The higher order
terms in eq.~\massshift\ vanish as the cutoff is removed
($R_0^{-1}\rightarrow \infty$). One can sum all the leading (in $N_c$)
non-analytic corrections to the baryon mass to a given order in the
chiral expansion, if one retains the corresponding terms in the
derivative expansion in the original Lagrangian, eq.~\aonelag.

We have shown above that the large-$N_c$ limit of QCD in the baryon
sector reduces to a classical field theory for mesons coupled to a
static source. The pion cloud has a hedgehog form when the baryon is
chosen to be in a state of definite collective coordinate $A$, rather
than a state of definite spin and isospin. In general, there will also
be a cloud of other mesons (such as the $\rho$) around the baryon, if
these mesons are included in the Lagrangian. Thus the large-$N_c$ limit
is exactly equivalent to a chiral bag picture of baryons
\chiralbag.\foot{The pion cloud couples to the source only at $r=R_0$
with the special choice of $J(r)$ in eq.~\jsource. In general, the pions
will couple over the entire region where the source is non-zero.}\ Note
that this identification depends on the $1/N_c$ expansion, but does not
require the chiral limit. We have also seen how the classical equations
correctly reproduce the order $N_c$ parts of the loop diagrams in the
standard chiral perturbation theory approach.

In general, the results we have obtained depend on the smearing radius
$R_0$ of the baryon source. This can be eliminated by taking
$R_0\rightarrow 0$ while at the same time including all the higher
derivative terms in the chiral lagrangian. For example, form factor
effects are included via such higher derivative terms. As $r\rightarrow
0$, the pion field $F(r)$ becomes stronger, and the higher order and
non-linear terms in the chiral lagrangian become important. It is also
necessary to renormalize the coupling $g_A$ in the Lagrangian
eq.~\pionb\ so that the physical pion-baryon coupling remains
finite.\foot{The physical pion-baryon coupling is determined by the
coefficient of $e^{-M_\pi r}/r$ in the pion tail at $r=\infty$. $g_A$ in
the Lagrangian must vanish as $R_0\rightarrow 0$ so that the physical
pion-baryon coupling remains finite.}\ The general solution of the full
non-linear theory (including higher derivative terms not included in
eq.~\aonelag) is complicated. It would be extremely interesting if one
could show that as $R_0\rightarrow 0$, $F(r)\rightarrow -\pi$ near
$r=0$. In that case, one could eliminate the central core since
$g_A^{\rm bare}\rightarrow 0$, and since the entire baryon number is
carried by the pion cloud \ref\goldstone{J.~Goldstone and R.L.~Jaffe,
\prl{51} {1983} {1518}}. The remaining object is a baryon whose
properties are completely determined by the meson Lagrangian---the
Skyrme soliton \ref\skyrme{T.H.R.~Skyrme, Proc.\ Roy.\ Soc.\ A260 (1961)
127}\anw. Whether this possibility holds is being investigated further.

\bigskip
\centerline{{\bf Acknowledgments}}
A recent paper by Dorey, Hughes and Mattis \ref\dhm{N.~Dorey, J.~Hughes, and
M.P.~Mattis, {\it Solvability, Consistency and the Renormalization Group in
Large-$N_c$ Models of Hadrons}, {\tt [hep-ph/9406406]}}\ discusses ideas
that are very closely related to those discussed here, and arives at
similar conclusions. I would like to thank M.P.~Mattis for discussing
their work prior to publication, and for comments on this manuscript. I
would also like to thank E.~Jenkins for helpful discussions.

This work was supported in part by the Department of Energy under grant number
DOE-FG03-90ER40546 and by a PYI award PHY-8958081 from the National Science
Foundation. I would like to thank the Aspen Center for Physics for hospitality
while this paper was being written.

\listrefs
\listfigs

\midinsert
\insertfig{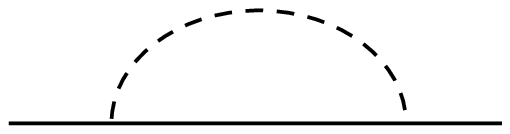}
\vskip-0.75in
\centerline{Figure 1}
\endinsert
\bye